\documentstyle[preprint,eqsecnum,aps]{revtex}
\begin{document}
\draft
\author{Mark A. Gunderson\thanks{Address after August 15: Department
of Physics, University of Florida, Gainesville FL 32611}
and Lars Gerhard Jensen\thanks{{\em E\/}-mail: ljensen@plains.nodak.edu}}
\address{Department of Physics, University of North Dakota, Grand Forks,
North Dakota 58202}
\date{\today}
\title{Boson Stars in the Brans-Dicke Gravitational Theories}
\maketitle
\begin{abstract}
Boson stars consist of a system of self-gravitating scalar fields which
form a macroscopic quantum state and are a possible dark matter candidate.
In this paper, we address the existence of boson stars in Brans-Dicke
gravity.  We show that solutions to the equations
of motion of a boson star in Brans-Dicke gravity exist for different
values of the coupling constant $\Lambda=\lambda M_{Pl}^2/4\pi m^2$,
where $\lambda$ is the quartic coupling, and $m$ the mass
of the boson field.
We find that there is only a small decrease of a few percent in the mass
of these objects when compared to similar solutions based on general
relativity.  We also consider solutions in the limit as $\Lambda
\rightarrow\infty$.
\end{abstract}

\narrowtext

\section{Introduction}

Alternate theories of gravity in general, and Brans-Dicke theories in
particular, have attracted considerable attention in recent years as
the framework for the study inflationary universe models~\cite{Infl}.
A main difference
between models of inflation based on ordinary gravity and models based
on Brans-Dicke gravity lies in the account of the completion of
the inflationary phase transition from the false vacuum phase (the de Sitter
phase) to the
present true vacuum phase (the Friedman-Robertson-Walker phase).
While the models based on ordinary
gravity require severe fine-tuning in order that the phase transition
can complete itself, the so-called {\em extended inflation} models based
on Brans-Dicke gravity explain the completion of the phase
transition in a more natural manner, which requires little or no fine-tuning:
The time-dependence of the gravitational constant causes the effective
cosmological constant in the de Sitter phase to decrease with time.
This eventually causes the expansion rate of the universe to drop below the
rate of formation of true vacuum bubbles, allowing the phase transition
to complete itself. This success, which first was pointed out by La and
Steinhardt, obviously provides good motivation for more detailed studies of
aspects of Brans-Dicke theories of gravity~\cite{LaSt}.

Most inflationary universe scenarios predicts the presence of
considerable amounts of dark matter --- typical estimates of the dark
matter abundance of the universe lie between 90\% and 99\% by mass.
While several candidates for the dark matter have been suggested, its
nature and existence is still an open question.
The candidates include neutrinos, several
exotic elementary particles, and a variety of compact objects such as
black holes, brown dwarfs, boson stars, etc. Our objective in this paper
is to study one of these --- the boson star --- in the context of
Brans-Dicke gravity.

Boson stars are compact stellar object composed of bosons which have
condensed into their ground state~\cite{Jetz}. They  were proposed by
Ruffini and Bonazzola, who first solved the equations of motion, and
demonstrated that such objects might indeed exist in
nature~\cite{RuBo}. In a later study Colpi, Shapiro and Wasserman were the
first to suggest that boson stars might contribute to the solution of the
dark matter problem~\cite{CSW}. In both of these studies the bosons are
described by
a scalar field coupled to ordinary Einstein gravity. In the simple
model considered by the former authors, the bosons are described by a
non-interacting, massive  scalar field. The authors assume that
the bosons are all in their ground state, and
then proceed to solve the coupled set of equations of
motion. Using a Schwarzschild metric to describe space-time, they find
the mass of the boson star to be of the order of $M\sim M^2_{pl}/m$,
where $m$ is the mass of the bose field. Even though this simple model
does open the interesting possibility that boson stars exists in nature,
it probably does not contribute to the
solution of the dark matter problem: For a
boson of mass $m\sim 1000$MeV one finds $M\sim 10^{-19}M_\odot$ for the
mass of the star. (Here $M_\odot$ is one solar mass.) Stars of this
small mass are not likely to contribute significantly to the dark matter
of the universe. The slightly
more realistic model studied by Colpi, Shapiro and
Wasserman includes self-interactions of the bosons. In their model the
bosons are described by a complex scalar field $\phi$ (to ensure that the
total charge of the field is conserved),
with a self-interaction of the form
$\frac{1}{4}\lambda |\phi|^4$. It is surprising that by including such
simple interactions the authors find stellar solutions of mass
$M\sim M^3_{Pl}/m^2$, in other words a mass of the same order of magnitude
as the Chandrasekhar mass. For a bosons of mass $m$ equal
to the proton
mass this corresponds to $M\sim M_\odot$. Thus, by allowing a
quartic interaction of the bosons, it is possible to
gain a factor of about
$10^{20}$ in mass, compared to a star formed from non-interacting bosons.
Such stars could contribute significantly to the dark matter
content of the universe.

Our aim in this paper is to re-analyze the work of reference~\cite{CSW}
in the context of the Brans-Dicke theories of gravity. In these
theories Newton's constant G has been replaced by a scalar field
$\phi_{BD}$, which is allowed to vary in space and time~\cite{BrDi,Wein}.
The question we pose here is which influence such a modification of
gravity has on the equations of motions of the boson star, and on the
solutions of these equations. In section II below we derive the
full set of coupled equations of motion of the star. In section III
we solve these numerically and compare our results to
earlier ones obtained by using Einstein's theory of gravity.

\section{Theory}

The action for the Brans-Dicke theory is
\widetext
\begin{equation}
I=\frac{1}{ 16\pi}\int d^4x\sqrt{-g}\left(-\phi_{BD} R -\omega g^{\mu\nu}
\frac{\partial_\mu\phi_{BD}\partial_\nu\phi_{BD}}{\phi_{BD}}\right)+
\int d^4x\sqrt{-g}{\cal L} \label{action}
\end{equation}
\narrowtext where $g= Det g_{\mu\nu}$.
The Brans-Dicke scalar field $\phi_{BD}$ plays
the r\^{o}le of Newton's
constant. The parameter $\omega$ is dimensionless and we take it
to be independent of $\phi_{BD}$. $R$ is the scalar curvature.
Einstein gravity can be recovered in the limit of
$\omega$ tending to infinity. Current observation places a lower limit
of 500 on the value of $\omega$~\cite{Reas}.
The matter content of the star is described by the Lagrangian density
$\cal L$ which we take to be that of a complex scalar field $\phi$.
It contains a quartic self-interaction
\begin{equation}
V(\phi)=\frac{1}{4}\lambda |\phi|^4
\end{equation}
and has the form,
\begin{equation}
{\cal{L}}=-\frac{1}{2}g^{\mu\nu}\partial_{\mu}\phi^{*}\partial_{\nu}\phi
-\frac{1}{2}m^2 |\phi|^2-\frac{1}{4}\lambda |\phi|^4
\end{equation}

Varying the action with respect to the metric and the scalar fields, we
obtain the following equations motion:
\widetext
\begin{mathletters}
\begin{eqnarray}
G_{\mu\nu} &=& \frac{8\pi}{\phi_{BD}}\; T_{\mu\nu} +\frac{\omega}
{\phi_{BD}^2}\left(\partial_{\mu}\phi_{BD}\partial_{\nu}\phi_{BD}
-\frac{1}{2}g_{\mu\nu}
\partial_{\lambda}\phi_{BD}\partial^{\lambda}\phi_{BD}\right)
\nonumber\\ &&
+\frac{1}{\phi_{BD}}(\phi_{BD\; ;\mu\; ;\nu} -
g_{\mu\nu}\phi_{BD\; ;\lambda}^{\;\;\;\;\;\;\;\;\;\; ;\lambda})\label{eq:einbr}
\\
0 &=& \phi_{;\lambda}^{\;\;\; ;\lambda}-m^2\phi-\lambda\phi^2\phi^{*}
\label{eq:sceq} \\
0 &=& \frac{2\omega}{\phi_{BD}}\phi_{BD\; ;\lambda}^{\;\;\;\;\;\;\;\;\;\;
;\lambda}-\omega\; \frac{\partial^{\lambda}\phi_{BD}\partial_{\lambda}\phi_
{BD}}{\phi_{BD}^2} + R \label{eq:sclfld}
\end{eqnarray}
\end{mathletters}
where $G_{\mu\nu}$ is the Einstein tensor. The energy-momentum
tensor $T_{\mu\nu}$ of the matter is given by
\begin{equation}
T_{\mu\nu}=\frac{1}{2}(\partial_{\mu}\phi^{*}\partial_
{\nu}\phi +\partial_{\mu}\phi\partial_{\nu}\phi^{*})-\frac{1}{2}g_{\mu\nu}
\left(g^{\rho\kappa}\partial_{\rho}\phi^{*}\partial_{\kappa}\phi +m^2 |\phi|^2
+\frac{1}{2}\lambda |\phi|^4\right) \label{eq:tuv}
\end{equation}
\narrowtext
If we substitute the trace of equation (\ref{eq:einbr})
\begin{equation}
R=-\frac{8\pi}{\phi_{BD}}T\frac{\omega}{\phi_{BD}^2}\partial^{\lambda}\phi_
{BD}\partial_{\lambda}\phi_{BD} +\frac{3}{\phi_{BD}}
\phi_{BD\  ;\lambda}^{\;\;\;\;\;\;\;\;\;\; ;\lambda}
\end{equation}
into equation (\ref{eq:sclfld}), we then find the following wave equation
for the Brans-Dicke scalar field,
\begin{equation}
\phi_{BD\; ;\lambda}^{\;\;\;\;\;\;\;\;\;\; ;\lambda}=\frac{8\pi}{2\omega
+3}T. \label{eq:brscfld}
\end{equation}

The star is a spherically symmetric system.  As such, space-time can
be described by the Schwarzschild metric, which is given by
\begin{equation}
d\tau^2 = -B(r)\; dt^2 + A(r)\; dr^2 + r^2\; d\Omega^2 \label{eq:schwar}
\end{equation}
thus,
\begin{eqnarray}
g_{00} &=& -B(r) \nonumber \\
g_{11} &=&A(r) \nonumber \\
g_{22}&=&r^2 \nonumber \\
g_{33}&=&r^2\sin^2 \theta \nonumber
\end{eqnarray}
and $g$ is given by
\begin{equation}
g = -A(r)B(r)r^4\sin^2\theta
\end{equation}

In this spherically symmetric geometry, we demand that the matter field
$\phi$ describing the bosons also be spherically symmetric.
This field has the form
\begin{equation}
\phi({\bf r},t)=\varphi(r)e^{-i\omega_{E}t} \label{eq:class}
\end{equation}
where $\varphi(r)$ is a real function and $\omega_{E}$ defines the
ground state energy
of the bosons.  (We could equally well consider ``antiboson stars'' with the
solution $\phi({\bf r},t)=\varphi(r)e^{i\omega_{E}t}$.)

Before we explicitly write out the equations of motion, let's introduce
the quantity ${\cal{M}}(x)$, which determines the amount of the
star's mass within the radius $x$, through the quantity
\begin{equation}
A(x)=\left(1-\frac{2{\cal{M}}(x)}{x}\right)^{-1}
\end{equation}
By using the Schwarzschild metric (\ref{eq:schwar}) along with the classical
field~(\ref{eq:class}) in the $(00)$ and $(11)$ field equations
(\ref{eq:einbr}),
the boson scalar wave equation (\ref{eq:sceq}), and the wave equation of
the Brans-Dicke scalar field (\ref{eq:brscfld}), then the equations of
motion for the boson star become
\widetext
\begin{mathletters}
\label{eq:motion}
\begin{eqnarray}
{\cal{M}}^\prime (x)&=&\frac{x^2}{2\Phi_{BD}}\left(\frac{2\omega +3}{2\omega
+4}\right)\left[\left(\frac{\Omega^2}{B}+1\right)\sigma^2 +\frac{\sigma^
{\prime\; 2}}{A} +\frac{\Lambda}{2}\sigma^4\right]+\frac{\omega x^2}{4A}
\left(\frac{\Phi_{BD}^\prime}{\Phi_{BD}}\right)^2 \nonumber \\
  & & + \frac{B^\prime x^2}{4AB}
\left(\frac{\Phi_{BD}^\prime}{\Phi_{BD}}\right)
  +\frac{x^2}{\Phi_{BD}}\left(\frac{1}{2\omega +4}\right)\left[\left(
 \frac{\Omega^2}{B} -2\right)\sigma^2-\frac{\sigma^{\prime\; 2}}{A}-
\Lambda\sigma^4\right] \label{eq:mbr}  \\
B^\prime (x) &=& \frac{AB}{x} - \frac{B}{x} + \frac{xAB}{\Phi_{BD}}\left(
\frac{2\omega +3}{2\omega +4}\right)\left[\left(\frac{\Omega^2}{B}-1\right)
\sigma^2+\frac{\sigma^{\prime\; 2}}{A}-\frac{\Lambda}{2}\sigma^4\right]
+\frac{\omega xB}{2}\left(\frac{\Phi_{BD}^\prime}{\Phi_{BD}}\right)^2
\nonumber \\
 & & +\frac{xB}{\Phi_{BD}}\left(\Phi_{BD}^{\prime\prime} -\frac{A^\prime}{2A}
 \Phi_{BD}^\prime\right)-
\frac{2xAB}{\Phi_{BD}}\left(\frac{1}{2\omega +4}\right)
 \left[\left(\frac{\Omega^2}{B}-2\right)\sigma^2 -\frac{\sigma^{\prime\; 2}}
 {A}-\Lambda\sigma^4\right] \label{eq:bbr}  \\
0 &=& \sigma^{\prime\prime}+\left(\frac{B^\prime}{2B}-\frac{A^\prime}{2A}
+\frac{2}{x}\right)\sigma^\prime+A\left[\left(\frac{\Omega^2}{B}
-1\right)\sigma - \Lambda\sigma^3\right] \label{eq:sc}  \\
0 &=& \Phi_{BD}^{\prime\prime}+\left(\frac{B^\prime}{2B}-\frac{A^\prime}{2A}
+\frac{2}{x}\right)\Phi_{BD}^\prime-\left(\frac{2A}{2\omega +4}\right)\left[
\left(\frac{\Omega^2}{B}-2\right)\sigma^2-\frac{\sigma^{\prime\; 2}}{A}
-\Lambda\sigma^4\right] \label{eq:scf}
\end{eqnarray}
\end{mathletters}
\narrowtext
The primes denote differentiation with respect to the re-scaled
radial coordinate $x$, and we have defined the dimensionless quantities
\begin{mathletters}
\begin{eqnarray}
x &=& mr \\
\Omega &=& \frac{\omega_{E}}{m} \\
\sigma &=& \sqrt{4\pi}\frac{\varphi}{M_{Pl}} \\
\Phi_{BD} &=& \left(\frac{2\omega +3}{2\omega +4}\right)\frac{\phi_{BD}}{M_
{Pl}^2} \label{eq:rescale} \\
\Lambda &=& \frac{\lambda}{4\pi}\left(\frac{M_{Pl}}{m}\right)^2
\end{eqnarray}
\end{mathletters}
where the parameter $\Lambda$ contains the scalar coupling $\lambda$, and
$M_{Pl}=\sqrt{\frac{\hbar c}{G}}$ defines the Planck mass.

Next, we address the boundary conditions that must be imposed on the
system (\ref{eq:motion}). In order to insure that gravitational
constant measured far away from the star is $G$, we must demand that,
$\phi_{BD}$ satisfies the following boundary condition (see Ref.~\cite{Wein}),
\begin{equation}\label{field}
\phi_{BD}(\infty)=\frac{1}{G}\left(\frac{2\omega +4}{2\omega +3}\right)
\end{equation}
With the re-scaling of equation~(\ref{eq:rescale}), the boundary
condition on the Brans-Dicke field takes the form
\begin{equation}
\Phi_{BD}(\infty)=1
\end{equation}
The boundary condition on the bose field at the enter of the star,
\begin{equation}
\sigma^\prime(0) =0
\end{equation}
follows from (\ref{eq:sc}) by demanding that the solution be regular
at $x=0$.  Similarly follows from (\ref{eq:scf}) that,
\begin{equation}
\Phi_{BD}^\prime(0) =0
\end{equation}
The remaining boundary conditions we impose on the system (\ref{eq:motion})
are
\begin{eqnarray*}
{\cal{M}}(0) &=& 0 \\
\sigma (0) &=& \sigma_{c} \\
\sigma^\prime (0) &=& 0 \\
B(\infty) &=& 1
\end{eqnarray*}
The value of the boson field at the center of the star $\sigma_{c}$
determines its central energy density.
The boundary condition on $B$ at
infinity is arbitrary and has here been chosen to unity.

Finally, before we proceed to present our results, we wish to point
out that in the limit of $\omega\rightarrow\infty$ the system of
equations of motion and boundary conditions above is in agreement
with general relativity:
Equation~(\ref{eq:brscfld}) implies that as $\omega$
becomes much larger than unity, then
\begin{equation}
\Phi_{BD}^\prime\sim O\left(\frac{1}{\omega}\right)
\end{equation}
Using this together with eq.~(\ref{field}) and taking the limit
$\omega\rightarrow\infty$, we recover the system of
equations of motion derived in Ref.~\cite{CSW} for general
relativity.

\section{Results}

In a similar manner as in the model presented by Colpi, Shapiro, and Wasserman,
we have sought to obtain non-singular, finite-mass, zero-node solutions to
the equations of motion of a boson star in the Brans-Dicke
gravitational theory.
With the currently accepted minimum value of $\omega=500$, the results
that we have obtained have shown that there is no
noticeable difference from the results based on general relativity.
For the purpose of developing a first understanding of the effects of a
variable gravitational constant on the boson star solutions, we will
here consider smaller values of $\omega$ also. Another justification
for this is that although $\omega$ must
exceed 500 today, there is no a priori reason why it could not have been
smaller in the past.\footnote{The possibility of a variable $\omega$ has
in fact already been considered in the so-called hyperextended models of
inflation~\cite{Stei}.}

In solving the equations of motion, we have used a Runge-Kutta method.
Figure~\ref{fig1} shows a plot of the results of our numerical
calculation of the mass of the star for $\omega=6$ given in units of
$M_{Pl}^2/m$,
\begin{equation}
M={\cal{M}}(\infty)\frac{M_{Pl}^2}{m}
\end{equation}
The mass is plotted as a function of $\sigma_{c}$ for different values
of $\Lambda$.

It is apparent from Figure~\ref{fig1} that for each value of $\Lambda$, we have
obtained a characteristic maximum mass for a specific value of $\sigma_{c}$.
The values of the maximum mass and its corresponding value of $\sigma_{c}$
for each value of $\Lambda$ along with the initial values of $B$ and
$\Phi_{BD}$ at the center of the star have been given in Table~\ref{tb1}.
In Table~\ref{tb2}, we have supplied the results that were calculated from the
general relativistic model presented by Colpi, Shapiro, and Wasserman.  When
the maximum masses obtained in the Brans-Dicke theory are compared with those
obtained in the general relativistic theory, a slight decrease in the
peak masses was found along with a shift of each of the peaks to a larger value
of $\sigma_{c}$ corresponding to a higher density at the center of the boson
star.  A similar effect was also seen in a study of neutron stars influenced by
Brans-Dicke gravity where it was shown that the form of the equation of state
determines how the mass will be affected~\cite{hillebrandt}.

In order to study this trend in a more quantitative manner, we plot the maximum
boson-star mass as a function of $\Lambda$ in Figure~\ref{fig2}.
Note that for $\Lambda\gg 1$, the maximum mass follows the
curve given by
\begin{equation}
M_{max} \approx 0.213\Lambda^{\frac{1}{2}}\frac{M_{Pl}^2}{m} \label{eq:M}
\end{equation}
Similarly, it was shown in Ref.~\cite{CSW} that for large values of $\Lambda$
the maximum mass followed the curve
\begin{equation}
M_{max} \approx 0.218\Lambda^{\frac{1}{2}}\frac{M_{Pl}^2}{m} \label{eq:bdma}
\end{equation}
When the maximum masses from the Brans-Dicke model are plotted as a function of
$\Lambda$, as in Figure~\ref{fig2}, they tend to follow the curve given by
equation~(\ref{eq:M}) and thus have a $\Lambda^{\frac{1}{2}}$
dependency for large values of $\Lambda$ just as in the general relativistic
model in reference~\cite{CSW}.

In order to explain the $\Lambda^{\frac{1}{2}}$ dependence of the
mass for large values of $\Lambda$, we first consider the plots of
$\sigma$ versus $x$ with $\sigma_{c} =0.06$ for $\Lambda =0$ and $\Lambda
=300$ in Figure~\ref{fig3} in a similar manner to that in Ref.~\cite{CSW}.
The curve for the large value of $\Lambda$ takes on a noticeably different
structure from the $\Lambda=0$ curve.  In this figure, we
see that for small $\Lambda$, $\sigma (x)$ falls off smoothly and eventually
exponentially to zero in a length $\sim \frac{1}{m}$.
However, for large $\Lambda$, we find that $\sigma (x)$ falls off rather
slowly out to about the radius
\[
\frac{\Lambda^\frac{1}{2}}{m}
\]\
with an exponential decline beyond this radius.  Because of the relationship
between $\sigma (x)$ and ${\cal{M}} (x)$, most of the mass in the large
$\Lambda$ case is found in the region of slow decline.  And, as $\Lambda$
increases, the region of slow decline becomes increasingly dominant, as can
be seen in Figure~\ref{fig3}.

These results suggest the same re-scaling of the equations of motion as seen
in coupling model based on general relativity, which was presented by Colpi,
Shapiro, and Wasserman.  However, the same type of re-scaling of the
Brans-Dicke field $\Phi_{BD}$ is not warranted as it does not exhibit the
same $\Lambda$ dependency as $\sigma (x)$.  Therefore, the resulting
equations of motion are still very difficult to solve analytically.

In Figure~\ref{fig3}, we compare the values of $\sigma (x)$ in the limit
of $\Lambda\rightarrow\infty$ with the exact values of $\sigma (x)$
for $\Lambda=300$ and $\sigma_{c}=0.06$.  Note that the approximate and exact
values agree except at radii larger than $\frac{\Lambda^{\frac{1}{2}}}{m}$.

The re-scaled mass as a function of the parameter $\frac{\Omega^2}{B(0)}$
with the additional boundary condition $\Phi_{BD} (\infty)=1$ is shown in
Figure~\ref{fig4}.
When these results are compared with the results of the same
approximation in the general relativistic theory, a slight decrease in the
maximum mass similar to the decrease seen in the exact values are observed.
The peak value of the mass in Figure~\ref{fig4} is
\begin{equation}
{\cal{M}}_{BD\; max} \approx 0.213
\end{equation}
and the peak is located at $\frac{\Omega^2}{B(0)}=0.312$.  This agrees with
the curve shown in Figure~\ref{fig2} and given by equation (\ref{eq:bdma}) for
large $\Lambda$.  The curve given by the general relativistic model in
Ref.~\cite{CSW} is given by
\begin{equation}
M_{max}=0.218\Lambda^{\frac{1}{2}}\frac{M_{Pl}^2}{m} \label{eq:grma}
\end{equation}

In general, for almost all $\Lambda$ and $\omega$,
the Brans-Dicke model of the boson star gives a maximum mass which
deviates only by a few percent or less from the
general relativistic model.  We find that as $\omega$ is
decreased the value of the maximum mass also decreases, and the
decrease becomes more dramatic as $\omega<10$. When $\omega$ is
large ($\omega\gg 50$), the maximum mass if found to be in exact
agreement with general relativity.

If the definition of $\Lambda$ is substituted into the maximum mass equation,
then
\[
M_{max}=0.06\lambda^{\frac{1}{2}}\frac{M_{Pl}^3}{m^2}
\]\
which is comparable to the Chandrasekhar mass for fermions of mass $\frac{m}
{\lambda^{\frac{1}{4}}}$.  Therefore, just as was observed in general
relativity, it is possible that fairly massive boson stars could have formed
in the early universe under the influence of Brans-Dicke gravity in the
presence of even a small coupling effect ($\lambda\ll 1$).

\section{Conclusion}

In this paper, we have analyzed the possible existence of boson star
solutions within the context of Brans-Dicke gravity.
We have demonstrated their existence
for the case of a star formed from bosons
interacting through a potential,
\[
V(\phi)=\frac{1}{4}\lambda |\phi|^4
\]\
When we
compare the masses of boson stars for $\omega=6$ in Brans-Dicke
gravity with similar solutions found by Colpi, Shapiro, and
Wasserman in the context of general relativity, we have found
a decrease of a few percent in the mass of the boson star.  However,
since these solutions have a mass on the order of the Chandrasekhar
mass, we have shown that it is possible for bosons stars to have a
significant contribution to the existence of dark matter in a universe
in which the force of gravity is determined by
a Brans-Dicke scalar field.

It should be noted that while the boson star solutions presented in
this paper corresponds to a value of $\omega=6$, solutions for
$\omega>6$ have also been found. These solutions show the same
features as the $\omega=6$ case, except that the maximal mass differ
even less from that of general relativity.

We finish by noting that one important aspect of these solutions
which has not been covered in this paper is their stability
under small fluctuations. The lack of analytical solutions makes
such a study very difficult --- even for the case of general
relativity no complete such analysis has been done. However
for the case of general relativity solutions which are stable under
radial fluctuations have been found~\cite{stab}. We expect
that a similar study of stability under radial fluctuations is
feasible in the context of Brans-Dicke gravity.

\acknowledgments

This work was supported in part by NSF grant \# EHR--9108770 to the
state of North Dakota. One of us (MG) also acknowledges support from
the North Dakota Space Grant Program, which is supported by the NASA
Space Grant Fellowship Program.

\textheight 550pt


%
%

\begin{figure}
\begin{center}
\setlength{\unitlength}{0.240900pt}
\ifx\plotpoint\undefined\newsavebox{\plotpoint}\fi
\sbox{\plotpoint}{\rule[-0.500pt]{1.000pt}{1.000pt}}%

\caption{This is a plot of the mass as a function of $\sigma_{c}$ for
different values of $\Lambda$ and $\omega=6$.}
\label{fig1}
\end{center}
\end{figure}

\begin{figure}
\begin{center}
\setlength{\unitlength}{0.240900pt}
\ifx\plotpoint\undefined\newsavebox{\plotpoint}\fi
\sbox{\plotpoint}{\rule[-0.500pt]{1.000pt}{1.000pt}}%
%
\caption{This is a plot of the maximum masses as a function of $\Lambda$
with $\omega=6$.
The solid curve corresponds to the approximate solutions and the
points correspond to the exact solutions seen in the previous figure.}
\label{fig2}
\end{center}
\end{figure}

\begin{figure}
\begin{center}
\setlength{\unitlength}{0.240900pt}
\ifx\plotpoint\undefined\newsavebox{\plotpoint}\fi
\sbox{\plotpoint}{\rule[-0.500pt]{1.000pt}{1.000pt}}%
\setlength{\unitlength}{0.240900pt}
\ifx\plotpoint\undefined\newsavebox{\plotpoint}\fi
%
\caption{This graph plots $\sigma$ versus the distance $x$ for
$\Lambda=0$ and $\Lambda=300$ and for $\sigma_{c}=0.06$ with $\omega=6$.
The points correspond to the results of the equations of motion modified
for large values of $\Lambda$.}
\label{fig3}
\end{center}
\end{figure}

\begin{figure}
\begin{center}
\setlength{\unitlength}{0.240900pt}
\ifx\plotpoint\undefined\newsavebox{\plotpoint}\fi
\sbox{\plotpoint}{\rule[-0.500pt]{1.000pt}{1.000pt}}%
%
\caption{This plot corresponds to the results of the mass found from
the modified equations of motion with $\omega=6$.}
\label{fig4}
\end{center}
\end{figure}

%
\end{center}
\end{table}

\begin{table}
\begin{center}
\caption{This table contains the maximum mass for each value of the coupling
$\Lambda$ from the general relativistic model along with the corresponding
values of $B(0)$ and $\sigma(0)$.}\label{tb2}
\begin{tabular}{cccc}
$\Lambda$ & $B(0)$ & $\sigma_{c}$ &
$M\left(\frac{m}{M_{Pl}^2}\right)$ \\ \hline
0 & 0.5020 & 0.275 & 0.633 \\
10 & 0.4341 & 0.225 & 0.921 \\
30 & 0.4358 & 0.150 & 1.334 \\
100 & 0.4225 & 0.090 & 2.254 \\
200 & 0.3900 & 0.070 & 3.131 \\
300 & 0.4066 & 0.055 & 3.814 \\
400 & 0.4296 & 0.045 & 4.379
\end{tabular}
\end{center}
\end{table}

\end{document}